\documentclass{bmcart}
\pdfoutput=1
\usepackage[utf8]{inputenc} 

\usepackage[pdftex,final]{graphicx}
\usepackage[final]{listings}
\definecolor{lightgray}{rgb}{.9,.9,.9}
\definecolor{darkgray}{rgb}{.4,.4,.4}
\definecolor{purple}{rgb}{0.65, 0.12, 0.82}
\lstdefinelanguage{JavaScript}{
  keywords={break, case, catch, continue, debugger, default, delete, do, else,
false, finally, for, function, if, in, instanceof, new, null, return, switch,
this, throw, true, try, typeof, var, void, while, with},
  morecomment=[l]{//},
  morecomment=[s]{/*}{*/},
  morestring=[b]',
  morestring=[b]",
  ndkeywords={class, export, boolean, throw, implements, import, this},
  keywordstyle=\color{blue}\bfseries,
  ndkeywordstyle=\color{darkgray}\bfseries,
  identifierstyle=\color{black},
  commentstyle=\color{purple}\ttfamily,
  stringstyle=\color{red}\ttfamily,
  sensitive=true
}

\startlocaldefs
\endlocaldefs

\begin{document}

\begin{frontmatter}

\begin{fmbox}
\dochead{Research}


\title{Open Chemistry: RESTful Web APIs, JSON, NWChem and the Modern Web Application}


\author[
   addressref={aff1},                   
   corref={aff1},                       
   noteref={n1},                        
   email={marcus.hanwell@kitware.com}   
]{\inits{MDH}\fnm{Marcus D} \snm{Hanwell}}
\author[
   addressref={aff2},
   noteref={n1},
   email={WAdeJong@lbl.gov}
]{\inits{BdJ}\fnm{Wibe A de} \snm{Jong}}
\author[
   addressref={aff1},
   email={chris.harris@kitware.com}
]{\inits{CJH}\fnm{Christopher J} \snm{Harris}}


\address[id=aff1]{
  \orgname{Kitware, Inc.}, 
  \street{28 Corporate Drive},                     %
  \postcode{12065}                                
  \city{Clifton Park, NY},                              
  \cny{USA}                                    
}
\address[id=aff2]{%
  \orgname{LBNL},
  \street{One Cyclotron Road},
  \postcode{94720}
  \city{Berkeley, CA},
  \cny{USA}
}


\begin{artnotes}
\note[id=n1]{Equal contributor} 
\end{artnotes}

\end{fmbox}


\begin{abstractbox}

\begin{abstract} 
An end-to-end platform for chemical science research has been developed that integrates data from computational and experimental approaches through a modern web-based interface. The platform offers a highly interactive visualization and analytics environment that functions well on mobile, laptop and desktop devices. It offers pragmatic solutions to ensure that large and complex data sets are more accessible. Existing desktop applications/frameworks were extended to integrate with high-performance computing (HPC) resources, and offer command-line tools to automate interaction---connecting distributed teams to this software platform on their own terms. The platform was developed openly, and all source code hosted on the GitHub platform with automated deployment possible using Ansible coupled with standard Ubuntu-based machine images deployed to cloud machines.

The platform is designed to enable teams to reap the benefits of the connected  web---going beyond what conventional search and analytics platforms offer in this area. It also has the goal of offering federated instances, that can be customized to the sites/research performed. Data gets stored using JSON, extending upon previous approaches using XML, building structures that support computational chemistry calculations. These structures were developed to make it easy to process data across different languages, and send data to a JavaScript web client.

\end{abstract}




\end{abstractbox}
%

\end{frontmatter}




\section*{Introduction}

The in-silico determination of chemical and materials properties is a vital capability that drives innovation across many market sectors. Its importance is reflected in the number of codes that can perform simulations over a broad range of levels of theory and length scale~\cite{nwchemsw, qchem, cp2k, abinit, wikipedia-qm}, and the enormous investments in experimental facilities that can also produce large data that is often difficult or impossible to reproduce. However, it is all too common for experimental and computational studies to take place independently, with large scale studies often involving heroic efforts developing one-off software projects dedicated to the specific resource, such as the Protein Data Bank~\cite{PDBarticle} (over 100,000 experimental structures), Materials Project~\cite{materialsProject} (over 33,000 simulated materials), and the Clean Energy Project~\cite{harvardCEP} (over 2.3 million calculated structures).

Driven by the U.S. Materials Genome Initiative the development of new and novel materials has become a multidisciplinary research endeavor where complex simulation and experimental data get integrated, and analytics such as machine learning techniques are utilized to aid in scientific discovery. This same multidisciplinary approach is becoming essential in chemical and biological research and development, in the design of new chemicals, biomolecules and drugs, or new energy efficient chemical production processes. 

There is a strong need to develop a collaborative scientific research software platform that enables researchers to define concepts and hypotheses, add them, and analyze integrated sets of experimental and computational data to offer effective knowledge discovery more universally. This goes well beyond a web portal to create a highly interactive platform integrating simulation, experimental data, and analytics, while leveraging semantic web technologies to support the federated storage of data across geographically dispersed sites.

The web has evolved significantly in the last two decades, and should be explored to assess what could be achieved in a platform that sought to use the latest open source tools, technologies, standards, and approaches to deliver an end-to-end platform for chemical/materials research. The basic approach employed was to develop a server component written in the Python language that exposes RESTful endpoints to interact with the data on the server. The Python code can use a number of existing core functionality, Python modules, and wrapped libraries developed in other programming languages.

Ideally no HTML, images, etc would be generated on the server, the server acts as a data server primarily through RESTful endpoints that accept/return data, along with user authentication (required by some endpoints). It can trigger calculations, perform analyses, and batch jobs in order to make the data discoverable. The RESTful server can then be consumed by a rich HTML5 web client, more traditional desktop clients, and from command-line clients or other servers to perform automated workflows.

In this work a rich HTML5 web interface was developed that made use of the server's RESTful API (application programming interface) briefly discussed above. It consumes data from the server, uploading/editing data, and maintaining local state in a one-page application reusing a popular open source HTML5 framework. The application integrated other open source frameworks for client side charting, and 3D rendering/visualization of molecular structure.

The web application was developed using open source tools, a number of frameworks, and was ``built'' as a static bundle of HTML5 web assets that are downloaded by the web client. It dynamically constructs the page in response to user interaction, server data, and other client events to provide a rich, interactive experience. This also meant that many interactions take place entirely on the client, requiring no interaction with or access to the server. This offers highly interactive data visualization and analysis, even on relatively low bandwidth links, once initial data about a molecule has been downloaded. 

\section*{Methods}

\begin{figure}[h!]
    \includegraphics[width=0.9\textwidth]{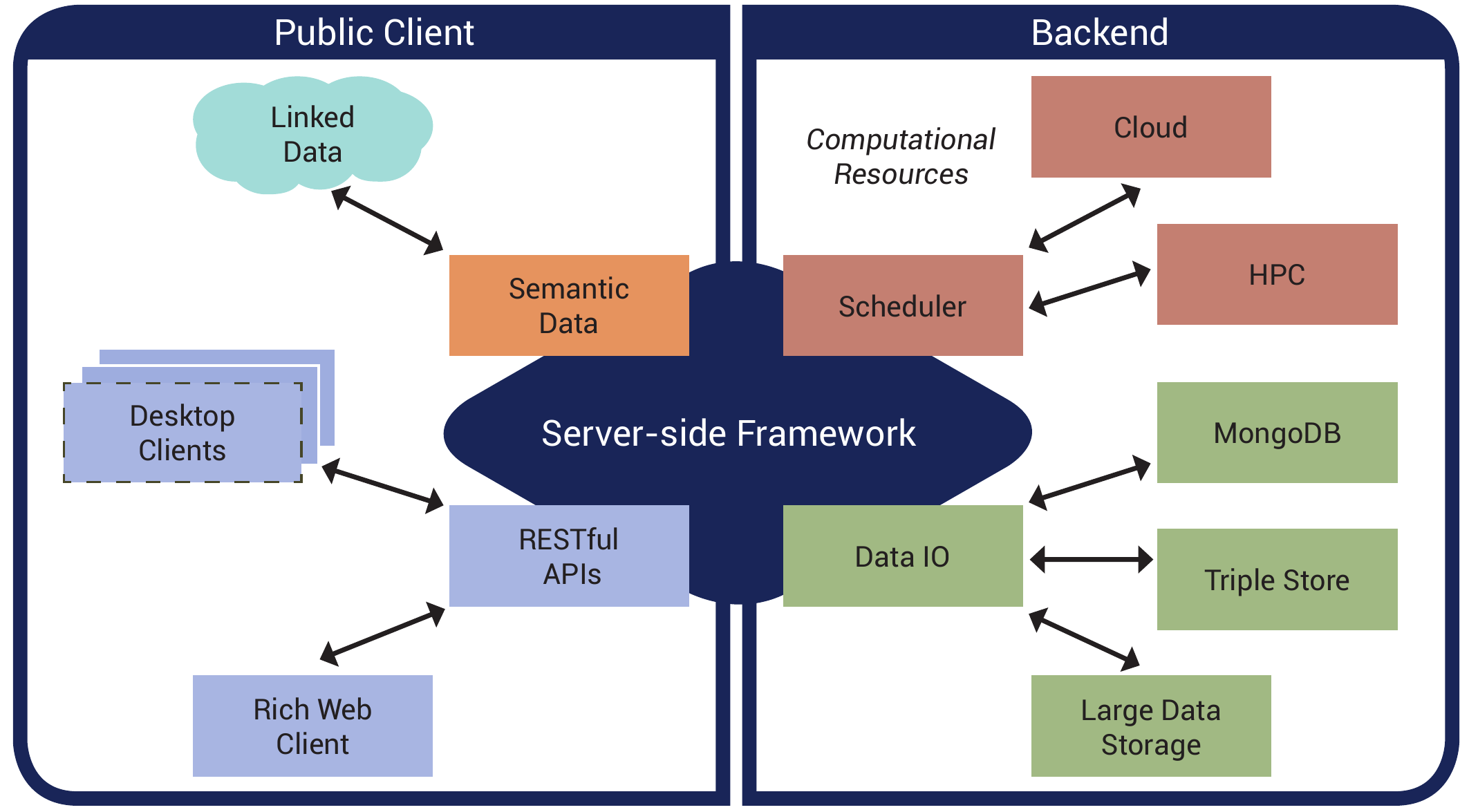}
    \caption{\csentence{Architecture of the chemical data platform.}
      Overview of the high-level architecture used for the chemical data platform.}
    \label{fig:architecture}
  \end{figure}

An open source prototype web platform was developed to demonstrate key capabilities in addressing the needs outlined in the introduction, and summarized in Figure~\ref{fig:architecture}. The application has a number of components developed in several languages following modern development methodologies. It was intentionally developed using some of the latest technology innovations, which means that it requires a modern web browser that supports WebGL in order to render 3D geometry, and it makes extensive use of HTML5. It is clear that not all devices/web browsers have full support for these technologies at this time, but that this support is already substantial and will grow in the coming years.

The server-side components were written in Python 3, with wrapped C++ code providing access to a number of existing chemistry/materials libraries such as Avogadro~\cite{avogadro} and Open Babel~\cite{openbabel}. The basis for the server-side project was the Girder project~\cite{girder}, which itself reuses a number of Python modules such as CherryPy that provides an extensible RESTful web server, Swagger to document the RESTful API, MongoDB to store data/metadata, and Virtuoso to store triples. The chemistry specific functionality was developed as additional API using the Girder project's plugin mechanism to add additional RESTful endpoints.

The client-side components were developed in modern HTML5, using open source web frameworks/technologies such as AngularJS 1.6 and Materials Design to provide a single page web application. 3DMol.js~\cite{3dmol} was used to render molecular geometry in 3D, D3~\cite{d3} to render charts, and responsive design elements to accommodate devices of various sizes/aspect rations. The capabilities developed have been demonstrated on desktop browsers, mobile phones, and tablets on the major operating systems. This includes Windows, macOS, Linux, iOS, and Android operating systems using browsers including Chrome, Firefox and Safari.

In addition to the significant developments made in the web platform, the reuse of the Avogadro 2 libraries enabled the rapid ingestion of chemical data. Extensions to the Avogadro 2 libraries, and several other components, were necessary. These changes have been merged into the main development branch, and were made available in the 1.90 release of the software. Significant capabilities added during the course of this project include visualization/animation of vibrational data, and additional file formats supporting the NWChem package. The JSON readers/writers built upon the JsonCpp library~\cite{jsoncpp}, and the capabilities were exposed to the Python-based server through Boost.Python wrapped calls to the C++ API.

The computational chemistry code used as a generator for the calculation results is the open source NWChem software suite~\cite{nwchemsw}. The JSON-Fortran library~\cite{json-fortran} was integrated into the NWChem source to enable the coded to write out a new JSON file in addition to the standard output or log file. APIs and interfaces between the Fortran-90 routines of the JSON-Fortran library and the Fortran-77 NWChem source were written to facilitate the transfer of the computational chemistry data into the JSON format. The full JSON enabled NWChem source code is available on Github~\cite{nwchem-json}. To enable the end-user to convert existing log files to the JSON format a Python 3 library was created. The library and examples are available in a separate Github repository~\cite{nwchem-json-convert}.

The collaborative nature of the development at Kitware and LBNL (Lawrence Berkeley National Laboratory) has created the opportunity to rapidly prototype new data structures, and look at the full end-to-end workflow from data generation, through to ingestion, analysis, and visualization. Python scripts were also developed to upload and add data files from the command-line, enabling ingestion of both existing data sets, and new ones as they are generated.

\section*{Results}

The software components were developed to serve the needs of data-centric chemistry research using open source approaches that embrace the use of open APIs, open data formats, and open components. The approach made extensive use of client-side rendering/interaction wherever practical, and focused on a server-side component that served data from RESTful endpoints using the web-native JSON format where possible. The development spanned a number of programming languages (Fortran, C++, Python, and JavaScript) in order to offer structured data that can be stored, queried, edited, and visualized.

\subsection*{MongoChemServer: Server Side Platform}

The server code was developed using the Python 3 language as a Girder plugin.
The Girder framework is an open source project led by Kitware, and released
under the Apache 2.0 license. It has three main components:
\begin{itemize}
 \item Data organization and dissemination
 \item User management and authentication
 \item Authorization management
\end{itemize}
It is developed as an extensible data management platform, which itself reuses a number of open source projects including CherryPy---``a pythonic, object-oriented web framework'', providing a solid foundation for the Girder platform. The code in the mongochemserver repository~\cite{mongochemserver} extends the functionality provided in a plugin that is loaded by the Girder process when it starts up. The plugin adds RESTful API, and reuses Girder core functionality and its core plugins for more generic features such as authentication, Gravatars, file upload/download, access control, etc.

The Girder platform provides deep integration with MongoDB, using that to store user credentials, access permissions, metadata, and other elements exposed via its plugin system. Among the most useful abstractions provided in the context of this project are the authentication, access permission, and file storage systems. Almost all of these concepts must be exposed on both the server and in the web client code to be used effectively.

The existing OAuth2 plugin was used, and coupled with Google's OAuth2 implementation to offer single-sign on. This can be replaced with other authentication schemes, or augmented with multiple options. For simplicity this was the only authentication scheme exposed in the prototype described, coupled with the use of encrypted SSL connections to provide secure authenticated access. This choice enabled the deployment of a demonstration to multiple locations, but was not always the most appropriate and will be augmented in future development to include integration with site-wide systems where appropriate.

\begin{figure}[h!]
    \includegraphics[width=0.8\textwidth]{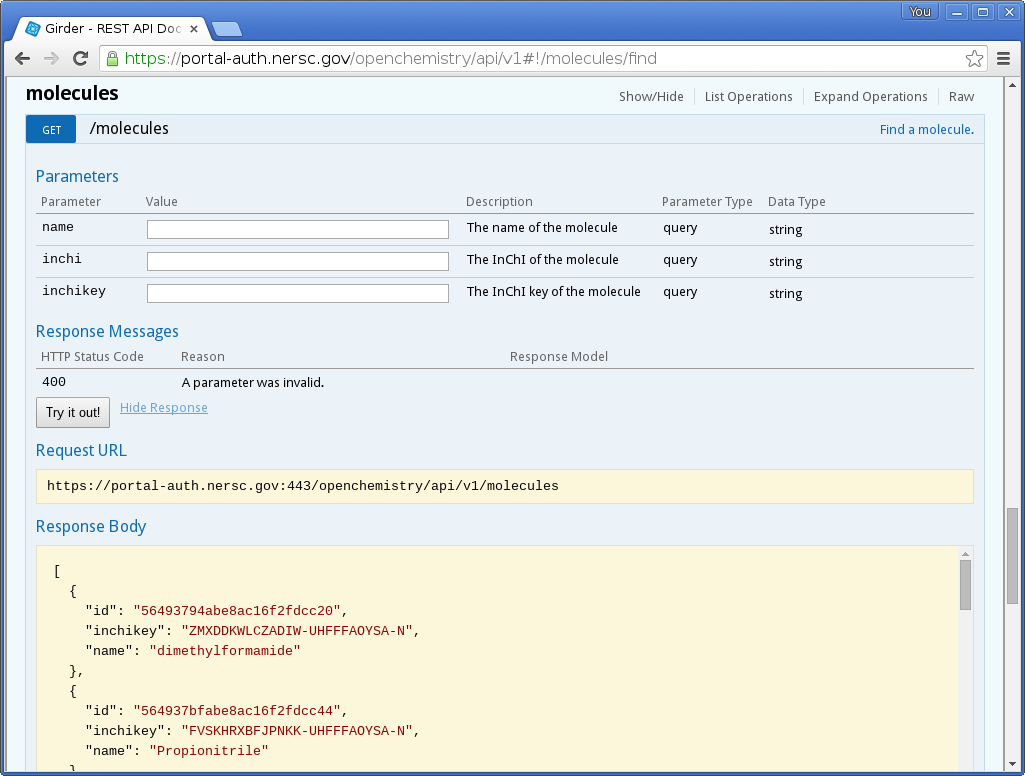}
    \caption{\csentence{Swagger documentation for RESTful API.}
      An example of Swagger being used to test some of the RESTful API in the `molecules' resource..}
    \label{fig:swagger}
  \end{figure}

The access permissions can be applied at several levels in the Girder code. A RESTful API must be exposed as a resource which resolves various paths, which refer to namespaces within the API prefix and are documented using a system called ``Swagger''~\cite{swagger}. This enables developers to document API as it is written, provides an HTML5 web client that exposes this documentation, and offers the ability to test API live on the web, shown in Figure~\ref{fig:swagger}. The API exposed uses decorators to express whether a given piece of API is public, or can only be accessed by authenticated users. API that requires authentication can apply further restrictions based on user privileges, and provide filtered results containing only data the authenticated user has the access privileges for.

File upload/download sounds quite simple on the surface, but it involves a number of distinct components in order to scale and integrate well in different environments. The Girder environment uses asset stores to abstract the storage backend, and the backend can then be mapped to file systems, S3 storage (as provided by Amazon EC2), and others. Large files must also be uploaded/downloaded in ``chunks'', something offered as part of the Girder file API and exposed in the client application. File system storage proved sufficient for the prototype, but future deployments would benefit from using large file stores, with extension to archive servers at supercomputing centers.

\subsection*{MongoChemClient: Rich HTML5 Web Client}

The Girder project has its own web interface, but this was not used---a custom user interface was developed in the mongochemclient repository~\cite{mongochemclient}. The web interface developed in the mongochemclient repository is a modern HTML5 web interface. This means that all HTML5 assets can be served as static files, and the page is built up dynamically on the client-side using the RESTful API to authenticate (if necessary), retrieve data, upload new date, and visualize data. A number of technologies and projects were leveraged in order to create a compelling, modern interface in a relatively short space of time.

  \begin{figure}[h!]
    \includegraphics[width=0.8\textwidth]{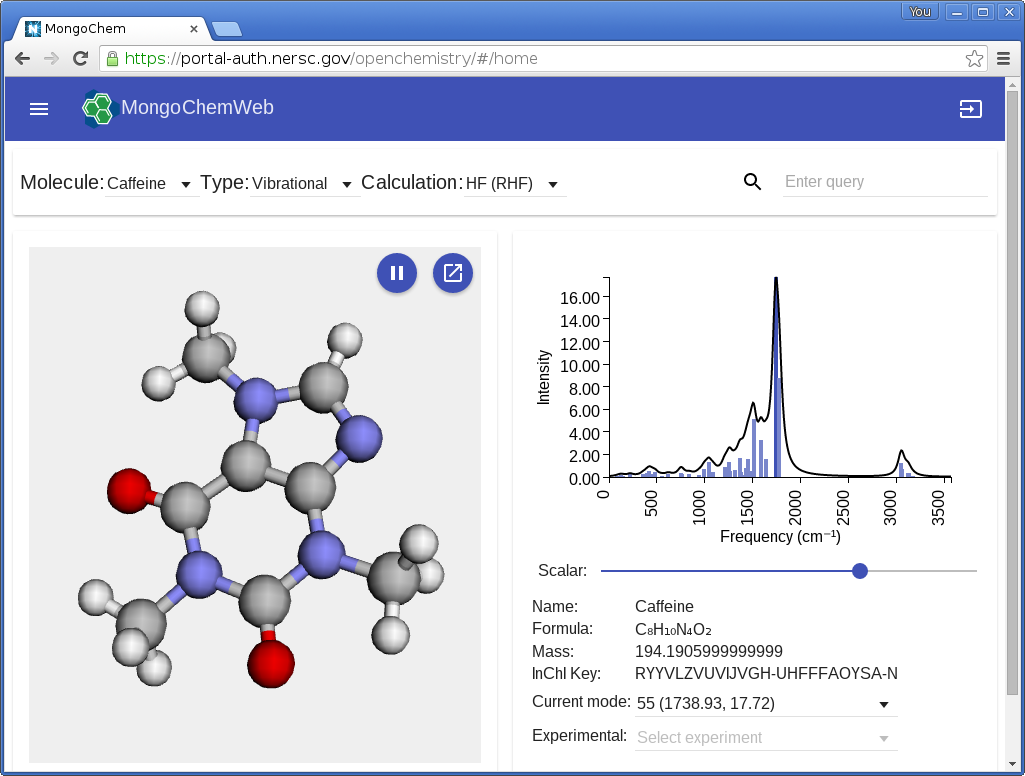}
    \caption{\csentence{HTML5 web client with molecular structure and vibrational modes.}
      The web client displaying a molecular structure in 3DMol.js, and selection of vibrational modes in a D3 chart.}
  \label{fig:webinterface}
  \end{figure}

The main framework used to manage interaction, react to events, and coordinate the single dynamic page approach was AngularJS. This framework is divided into a number of modules that provide various services/extensions, such as easy access to RESTful APIs, animations, routing (where the URL is updated to reflect the current `location' (state) despite being in a single-page web application), and overall look and feel (Materials design in this instance). AngularJS was chosen for the rich feature set, maturity, and encapsulation of components along with its powerful web page layout framework. An example of the single page application in action is shown in Figure~\ref{fig:webinterface}, where a molecular structure can be seen beside a plot of vibrational modes that can be animated in the 3DMol.js based geometry viewer.

Instead of writing HTML code directly the Jade language was chosen for its expressiveness, and conciseness. The translation to HTML is rather simple, and it offers a number of conveniences over directly using HTML. The same reasoning led the team to choose Stylus for CSS code, leading to much more concise code in a more expressive form. The Jade and Stylus languages are both well supported within webpack, which was used as both a development tool when working on new code, and as a static asset compiler when deploying code to production machines. Beyond these capabilities, Node.js also has a robust package management system, offering seamless installation/deployment of AngularJS, and even git repositories that contain other web-based components.

The static HTML5 content that serves as the frontend, and the dynamic RESTful APIs offering access to the data must be presented to the web client. This is where the NGINX web server~\cite{nginx} came in, offering an SSL-enabled endpoint for encryption, serving the static content in the web root, and proxying requests to the /api/v1 prefix to the Python-based backend. The Python-based backend also had access to the MongoDB server where all metadata, access controls, and links to files were stored. There is also an asset store, where a simple on-disk asset store was used.

This describes the very basic layout of the prototype platform, and some of the details of how this was extended will be detailed in the following text. It is clear that even the early prototype involves a number of build/deployment systems, and that the deployment repository was necessary to coordinate the task of deploying everything to the right location, with compatible software versions, and ensuring services are brought up/down in the correct order.

\subsection*{JSON Data Formats}

A JSON format for chemical data from computational chemistry simulations has been in development for a number of years, starting life within the Avogadro 2 project as a simple means of data storage and exchange. The JSON format was developed based upon the requirements of storing data in BSON using MongoDB, communicating chemical data over JSON-RPC 2.0 between desktop applications, and later using web services. It was also motivated by the need for a format to support an application being developed to edit and communicate chemical structures.

The format was tested with small molecules through to molecules with millions of atoms using a philosophy of being machine readable, efficient, and avoiding repetition. It was written to represent a simple  mapping of the in memory data structures. It is based on many of the concepts developed in the CML format~\cite{CML} as part several ongoing collaborations under the broad umbrella of the Blue Obelisk~\cite{blueobelisk, blueobelisk2}. We chose JSON because it is less cumbersome than XML, with simpler parsers, and is a native format of the web.

This JSON format exposed the internal data model used in Avogadro 2, and was used as the transport layer to the 3DMol.js client-side rendering. It was extended to expose vibrational modes, and a primitive container for volumetric data already present in the 3DMol.js project. The current state of the Chemical JSON format is documented in a Github repository~\cite{chemicaljson-git} and will continue to evolve as it is extended to serve more use cases.

A second, more comprehensive ExtendedChem JSON format was developed with the goal to encapsulate the data from computational chemistry software, such as the open-source NWChem. The goals of the two formats were somewhat orthogonal. While ExtendedChem JSON expresses the properties of a single molecule in machine readable arrays with all data that might be visualized/displayed, the Chemistry format is focused on storing all the essential input and output data from the computational chemistry software simulation (here for NWChem) for the (possibly multiple) jobs that were run as the calculation proceeded. This is somewhat akin to the typical molecular file formats (XYZ, CML, SDF, etc) versus a more structured quantum mechanical log file---a snapshot of definite structure versus data discovery and recording.

This new format focuses on satisfying the need to replace a log file with structured output, offering output of multiple calculations as a computational chemistry code like NWChem is executed. The JSON file generated by the NWChem application builds upon previous work done integrating CML into NWChem~\cite{nwchemAvogadro} (an XML-based format), and is designed around the notion of using objects as the primary representation. As a starting point the CML naming and conventions were adopted and extended with new concepts. The format was developed and designed to be portable to multiple computational chemistry codes but in this work firmly targeted the NWChem code.

The relevant section representing the geometry/structure of JSON files generated in the two formats, in this case for a water molecule, are compared to clearly show the difference in design philosophies, i.e. arrays vs objects. The listing below shows a molecule block within the Chemical JSON format:

\begin{lstlisting}
{
  "chemical json": 0,
  "atoms": {
    "coords": {
      "3d": [ 0.00, 0.00, 0.14,
             -0.76, 0.00,-0.46,
              0.76, 0.00,-0.46 ]
    },
    "elements": {
      "number": [8, 1, 1]
    }
  },
  "bonds": {
    "connections": {
      "index": [0, 1, 0, 2]
    },
    "order": [1, 1]
  }
}
\end{lstlisting}

The format focuses on using multiple arrays to store different aspects of atoms, bonds, etc---such as the atomic number, bond order, etc. The 3D coordinates are in a single array, and each 3D vector is offset by 3N where N is the index of the atom. This offers simplicity, compactness, and a simple representation with efficient storage of atomic data. It does not provide a layout that is focused on human readability, and it is not intended to be used as a directly editable set of objects in Python/JavaScript without some code in front of it to aid in manipulation/keeping the representation consistent.

The listing below shows the same molecule block within the ExtendedChem JSON format:

\begin{lstlisting}
{
  "molecule": {
    "id": "Molecule.1",
    "atoms": [
      {
        "id": "Atom.1.Mol.1",
        "elementLabel": "o",
        "elementSymbol": "O",
        "elementNumber": 8,
        "elementName": "Oxygen",
        "cartesianCoordinates": {
          "value": [
            0.0E+0,
            0.0E+0,
            0.2106360400000002E+0
          ],
          "units": "bohr"
        }
      },
      {
        "id": "Atom.2.Mol.1",
        "elementLabel": "h",
        "elementSymbol": "H",
        "elementNumber": 1,
        "elementName": "Hydrogen",
        "cartesianCoordinates": {
          "value": [
            -0.1841188380000002E+1,
            0.0E+0,
            -0.8425441600000008E+0
          ],
          "units": "bohr"
        }
      },
      {
        "id": "Atom.3.Mol.1",
        "elementLabel": "h",
        "elementSymbol": "H",
        "elementNumber": 1,
        "elementName": "Hydrogen",
        "cartesianCoordinates": {
          "value": [
            0.1841188380000002E+1,
            0.0E+0,
            -0.8425441600000008E+0
          ],
          "units": "bohr"
        }
      }
    ]
  }
}
\end{lstlisting}

In contrast to the Chemical JSON listing, the information of each atom in ExtendedChem JSON is collected in a single object. A key aspect of this format is the explicit definition of units, a key aspect brought in for the CML specification. It is relatively simple to map between the two formats, and this is now possible in the Avogadro 2 libraries which feature readers for both formats.

In developing the ExtendedChem JSON format, some additional referencing features were introduced that are not natural to JSON. The listing below  shows an excerpt of a JSON data file containing the calculation setup and some of the properties that can be calculated with quantum chemistry software:

\begin{lstlisting}
      {
        "calculationType": "molecularProperties",
        "molecularFormula": "H2O",
        "id": "calculation.4",
        "calculationSetup": {
          "outputVectors": "./prop_h2o_run.movecs",
          "molecularSpinMultiplicity": 1,
          "molecule": "Molecule.2",
          "charge": 0,
          "waveFunctionType": "RHF",
          "numberOfElectrons": 10,
          "basisSet": "BasisSet.1",
          "waveFunctionTheory": "Hartree-Fock",
          "inputVectors": "./prop_h2o_run.movecs",
          "id": "calculationSetup.4"
        },
        "calculationResults": {
          "molecularProperties": [
            {
              "Molecule": "Molecule.2",
              "dipoleMoment": {
                "totalMoment": {
                  "units": "atomic units",
                  "value": 0.8052087008
                },
              },
              "quadrupoleMoment": {
                "diamagneticSusceptibility": {
                  "units": "atomic units",
                  "value": 18.596483
                },
                "momentXZ": {
                  "units": "atomic units",
                  "value": 0.0
                }
              }
            },
            {
              "atom": "Atom.1.Mol.2",
              "diamagneticShielding": {
                "units": "atomic units",
                "value": 23.457292
              }
            }
          ]
        }
      }
\end{lstlisting}

Input structures in computational chemistry codes are set up to allow a sequence of calculations to be executed in one single run. To store this sequence, the ExtendedChem JSON format has adopted the CML approach of storing each calculation in a ``calculations'' array. The listing above shows an excerpt of the  fourth  calculation in this series. Often, data from previous calculation steps in the sequence is reused in the quantum chemistry code. To avoid duplicating information, ``id'' tags were used to provide a pointer to the first mention of the JSON object.  An example of this in the listing above are the key ``molecule'' to ``Molecule.2'', which is defined as the id-tag in the molecule block in the previous listing. Another example is the ``basisSet'' keyword in the ``calculationSetup'' block. Even the ``calculationSetup'' could be the same for multiple calculations, and could be pointed to in this fashion. The same approach is used to identify the atom within the molecule to which the calculated atomic properties belong. Here the ``atom'' keyword points to atom ``Atom.1.Mol.2'' in the earlier ExtendedChem JSON listing for the molecular geometry. Essentially, this referencing approach mimics the ability to link different sets of data. The referencing feature is missing in the current JSON specification but native to XML. The JSON pointer has been proposed in RFC 6091~\cite{RFC-6091} and a draft RFC is being developed for the JSON reference~\cite{RFC-reference}. While JSON reference would provide the necessary flexibility, either could  readily be adopted in the format in the near future.

Examples of complete ExtendedChem JSON  files generated by the NWChem quantum chemistry code can be found on  the Github repository~\cite{nwchem-json-convert}. 

The JSON file can be submitted to the web platform developed, and relevant metadata will be extracted. The format is capable of encapsulating jobs that have multiple steps, and it has been demonstrated with vibrational and electronic structure data. The format breaks most elements of the output into distinct objects, and it has been designed to be more semantically expressive than the Chemical JSON format developed as part of the Avogadro 2 project. 

The two formats were used as part of this project, with several extensions made to the Chemical JSON format in order to support data needed by the project. They are both supported by the Avogadro 2 libraries, and exposed in the Python-wrapped API used on the server side. The Chemical JSON format is designed in a pragmatic fashion, making extensive use of arrays, along with key-value pairs for simple properties. The format closely matches in-memory structures, it is optimized for storage in BSON (the binary form of JSON used internally by MongoDB), is easy to visualize and store as a single document in MongoDB. The new ExtendedChem JSON format developed tends to represent everything as an object, can store multiple configurations/states of a molecule, and uses links to reuse objects that have not changed. This makes it more complex to parse, but more expressive, and it more closely reflects what is currently stored in log files.

This collaborative development has highlighted extensions needed in the internal models used in the Avogadro 2 projects, the 3Dmol.js project, and in the model employed in the web framework. Ultimately, any given visualization can only show a single state, and the web and desktop frameworks both generally assume a file only has a single state with minimal links back to the input file. This concept will need to be extended in order to visualize and analyze data from more complex multi-step files, and some of this development was started in the work described.

In the future we would like to extend this approach to use and draw upon new developments in JSON-LD~\cite{jsonld} that aims to bring linked data and meaning to JSON using the same approaches used in semantic data structures.

\section*{Discussion}

Research in chemistry is becoming more data intensive, and as a result it is vital that we act to create platforms that can be easily deployed, used, and shared with a focus on data. This requires standard formats that support a multitude of facets of chemical data that leverage industry standard technologies such as XML, JSON, and the semantic web.

As we move closer to a landscape where the web is an essential part of any workflow we must embrace web-native formats, generally using JSON and similar containers, to enable the free exchange, analysis and visualization of data. This can be further augmented through the use of JSON-LD, and is something the authors would like to do in order to offer semantic meaning and enable the simple transformation of JSON-LD documents to triples.

Once these components are available the open source code referenced here, and available as a demonstration capability, show what can be built around it. The RESTful APIs, databases, and HTML5 frontends can be readily coupled with command line and powerful desktop applications to offer a compelling ecosystem of data-centric applications that serve the chemical research enterprise.

The definition of suitable schema, codification of that into JSON-LD, and community agreement still represent significant barriers to wider adoption. The development of permissively licensed open source code, which can be readily inspected, reused, and modified is one approach to improving the situation. Coupled with the use of community platforms such as GitHub for collaborative development, and workshops gathering interested experts, these approaches can be iterated upon and standardized.

\section*{Conclusions}

A ``full-stack'' open source web application was developed, making heavy use of Python and open source community tools on the backend, and HTML5/JavaScript on the frontend. Years of experience in using, converting, and developing data formats was drawn upon to create some new formats for the output of structured data from the NWChem code, ingestion into databases, and the subsequent visualization/analysis of data. This was demonstrated in a new web client, and added to the Avogadro 2 desktop application. The primary data types were 3D chemical structure, electronic structure, and vibrational modes on the web and desktop.

If the field is to progress it is important to move away from disparate formats developed using non-standard line printer style formats. The ingestion and normalization of data still represents a serious challenge to reaping the benefits powerful data-centric platforms promise. Offering simple, well documented formats, and a powerful, open, reference platform is one path to show the benefits, and begin working with the wider community to share data more widely while encouraging greater reuse.


\begin{backmatter}

\section*{Declarations}

\subsection*{Ethics approval and consent to participate}

Not applicable.

\subsection*{Consent for publication}

Not applicable.

\subsubsection*{List of abbreviations}

3D---Three dimensional\\
API---Application programming interface\\
CSS---Cascading style sheets\\
CML---Chemical markup language\\
HTML---Hypertext markup language\\
HPC---high-performance computing\\
JSON---JavaScript object notation\\
LBNL---Lawrence Berkeley National Laboratory\\
REST---Representational state transfer\\
SSL---Secure sockets layer\\
XML---Extensible markup language

\subsubsection*{Availability of data and materials}

All code and data discussed in this manuscript is available under 
open licenses on GitHub. The following repositories contain all the 
data and materials:

\begin{itemize}
 \item Chemical data RESTful server~\cite{girder,mongochemserver}
 \item Web frontend single-page application~\cite{mongochemclient}
 \item Chemical JSON specification and examples~\cite{chemicaljson-git}
 \item NWChem with JSON output~\cite{nwchem-json}
 \item Conversion tool from NWChem log files to JSON~\cite{nwchem-json-convert}
\end{itemize}

\subsection*{Competing interests}

The authors declare that they have no competing interests.

\subsection*{Funding}

The authors would like to acknowledge funding from the Department of Energy's Office of Science, under contract DE-SC0013250.

\subsection*{Author's contributions}

MDH and CJH developed the client and server code, new file format support in Avogadro 2, and the Chemical JSON file format. WDJ developed the CompChem JSON data format, the JSON writer in the NWChem application and the Python NWChem output converter, and generated the simulation data used in testing. MDH and WDJ wrote the manuscript. All authors read and approved the final manuscript.

\subsection*{Acknowledgements}

Not applicable.


\bibliographystyle{bmc-mathphys} 
\bibliography{bmc_article}      







\end{backmatter}
\end{document}